\documentclass[letterpaper,titlepage,11pt]{article}

\pdfoutput=1

\usepackage{amssymb,amsmath,amsfonts, graphics}
\usepackage{epsfig}
\usepackage{ccaption}
\usepackage{graphicx}

\usepackage[
      colorlinks=true,
      linkcolor=blue,
      urlcolor=blue,
      filecolor=black,
      citecolor=red,
      pdfstartview=FitV,
      pdftitle={},
        pdfauthor={Tomas Andrade, Roberto Emparan, David Licht, Raimon Luna},
        pdfsubject={},
        pdfkeywords={},
        pdfpagemode=None,
        bookmarksopen=true
      ]{hyperref}

\def\clock{{\count0=\time
           \divide\count0 60
           \ifnum\count0<10 0\fi\the\count0
           \multiply\count0 -60 \advance\count0 \time
           :\ifnum\count0<10 0\fi \the\count0
         }}
\newcommand{\timestamp}{{\small\vbox{\hbox{\tt\jobname.tex}
\hbox{\the\day/\the\month/\the\year, \clock}}}}

\newcommand{\beq}{\begin{equation}}
\newcommand{\eeq}{\end{equation}}
\newcommand{\beqa}{\begin{eqnarray}}
\newcommand{\eeqa}{\end{eqnarray}}
\newcommand{\bea}{\begin{eqnarray}}
\newcommand{\eea}{\end{eqnarray}}

\newcommand{\ie}{{\it i.e.,\,}}
\newcommand{\eg}{{\it e.g.,\,}}

\newcommand{\lp}{\left(}
\newcommand{\rp}{\right)}

\newcommand{\ord}[1]{{\mathcal O}\lp #1\rp}

\def\v2#1{{\color{blue}{#1}}}

\begin{document}

\begin{titlepage}
\leftline{}
\vskip 2cm
\centerline{\LARGE \bf Cosmic censorship violation in}
\bigskip
\centerline{\LARGE \bf black hole collisions in higher dimensions} 
\vskip 1.6cm
\centerline{\bf Tom{\'a}s Andrade$^{a}$, Roberto Emparan$^{a,b}$, David Licht$^{a}$, Raimon Luna$^{a}$}
\vskip 0.5cm
\centerline{\sl $^{a}$Departament de F{\'\i}sica Qu\`antica i Astrof\'{\i}sica, Institut de
Ci\`encies del Cosmos,}
\centerline{\sl  Universitat de
Barcelona, Mart\'{\i} i Franqu\`es 1, E-08028 Barcelona, Spain}
\smallskip
\centerline{\sl $^{b}$Instituci\'o Catalana de Recerca i Estudis
Avan\c cats (ICREA)}
\centerline{\sl Passeig Llu\'{\i}s Companys 23, E-08010 Barcelona, Spain}
\smallskip
\vskip 0.5cm
\centerline{\small\tt tandrade@icc.ub.edu,\, emparan@ub.edu,} 
\smallskip
\centerline{\small\tt david.licht@icc.ub.edu,\, raimonluna@icc.ub.edu}

\vskip 1.6cm
\centerline{\bf Abstract} \vskip 0.2cm \noindent
We argue that cosmic censorship is violated in the collision of two black holes in high spacetime dimension $D$ when the initial total angular momentum is sufficiently large. The two black holes merge and form an unstable bar-like horizon, which grows a neck in its middle that pinches down with diverging curvature. When $D$ is large, the emission of gravitational radiation is strongly suppressed and cannot spin down the system to a stable rotating black hole before the neck grows. The phenomenon is demonstrated using simple numerical simulations of the effective theory in the $1/D$ expansion. We propose that, even though cosmic censorship is violated, the loss of predictability is small
independently of $D$. 

\end{titlepage}
\pagestyle{empty}
\small
\normalsize
\newpage
\pagestyle{plain}
\setcounter{page}{1}



\noindent\textbf{Introduction.} The cosmic censorship (CC) conjecture \cite{Penrose:1969pc} raises the question of whether classical gravitational dynamics can drive a low-energy configuration into an accessible regime of quantum gravity, with Planck-scale curvatures and energy densities visible by distant observers. There is by now convincing evidence of processes that evolve towards CC violation (\eg \cite{Choptuik:1992jv,Lehner:2010pn,Figueras:2017zwa}), but the circumstances can vary and important details often remain to be ascertained: are CC violations generic, or do they instead require initial fine-tuning? What is the size of the violation, \ie\ the fraction of the initial energy that goes into Planckian densities, and how strong is the loss of classical predictability? Are there any constraints on, or implications for, these violations from thermodynamics and holographic entropy bounds?

In the investigation of this problem it is useful to explore a wide range of gravitational dynamics. 
The addition of spatial dimensions reveals suggestive paths towards the quantum-gravitational regime: Einstein's theory in  dimensions $D>4$ contains black strings whose horizons, resembling fluid jets that break up into droplets, are unstable to growing inhomogeneities along their length, with necks pinching where spacetime curvature diverges \cite{Gregory:1993vy,Lehner:2010pn}. 

In this article we present evidence of a violation of CC closely related to the one in unstable black strings, but which appears naturally in the quintessential phenomenon of General Relativity: the collision and merger of two black holes. It only requires a sufficient total angular momentum in a collision in high enough $D$.

The physical picture can be easily understood. When the two black holes collide, they coalesce forming a single, dynamically evolving horizon. For low total angular momenta the system settles into a stationary state of an axisymmetric rotating black hole. However, if the angular momentum is larger, the transient horizon is elongated, in the shape of a rotating bar or dumbbell. Beyond a certain value of the spin, this configuration is unstable in a manner similar to the instability that drives black strings towards a CC-violating break up \cite{Andrade:2018nsz}, \ie the bar forms a neck in its middle that pinches down with diverging curvature. Moreover, the emission of gravitational radiation from the rotating bar, which might spin it down to a stable rotating black hole, is suppressed when $D$ is large \cite{Cardoso:2002pa,Emparan:2013moa}. Therefore, the formation of the neck dominates the evolution.

It is a plausible speculation that the fate of the high-curvature neck is controlled by the same physics that governs the endpoint of black hole evaporation---that is, when the surface gravity at the pinch reaches the Planck scale, it evaporates by emitting Planck-energy quanta (much like fluid jets break up by evaporating necks with high surface tension). Here we expect that the neck in the bar snaps, leaving two separate black holes that fly apart from each other.

The brief instant of break up is the only moment when quantum gravitational effects enter. However, the CC-violating region is small, Planck-sized, and only a few Planckian quanta can be present. Therefore, (assuming that quantum gravity does indeed effect the break up) the classical evolution will resume after a few Planck times with hardly a loss of predictability: the trajectories of the ejected black holes can be determined, up to errors that are small in the ratio of the Planck mass to the black hole mass, by the state of the system up to the moment of break up. If this picture is correct, then, even if CC is violated, its spirit remains unchallenged: classical relativity describes the physics seen by observers outside the black holes accurately, with only minimal quantum input that does not entail macroscopic disruptions.

We will demonstrate the phenomenon just described in black hole collisions in a large enough number of dimensions $D$ (necessarily larger than four, as we will see) using the $1/D$ expansion within the framework recently developed in \cite{Andrade:2018nsz}. The importance of these calculations is that, although they do not allow us to capture the moment when the singularity appears (a regulator is always present), nor its detailed structure, they do show that (a) an elongated horizon forms in the collision, and (b) this configuration grows a neck and pinches in its middle. That is, the system is driven to a situation where it has been shown that a pinch of this kind leads to diverging naked curvature \cite{Lehner:2010pn,Emparan:2018bmi}.
In a forthcoming article we will present further details and extensions of the calculations discussed here \cite{longpaper}.

\medskip
\noindent\textbf{Method.} In the approach of \cite{Andrade:2018nsz} to the physics of large-$D$ black holes, these appear as spheroidal blobs on the horizon of a thin black brane. These blobs not only account correctly for the mass, entropy, and angular momentum of stationary Myers-Perry (MP) black holes in the limit $D\to\infty$: they can also be set in linear motion by a boost, and when perturbed, their vibrations reproduce with accuracy the quasinormal modes of the black hole, axisymmetric or not. The presence of the thin black brane does not affect these properties of single black holes, but acts as a regulator when two black holes either touch or split apart: the horizons never actually merge nor break up, but are always continuously joined by the thin black brane. This feature allows us to follow the entire evolution of the system. 

These black-hole blobs evolve according to the equations for the large-$D$ effective dynamics of a neutral black brane. As shown in \cite{Emparan:2015gva,Emparan:2016sjk}, any solution of
%
%
%
\beq
	\partial_t m+\nabla_i(mv^i) =0, \qquad
	\partial_t(mv^i)+\nabla_j \tau^{ij} =0\label{eqs}\,,
\eeq
where
\begin{equation}
	\tau_{ij} = m (v_i v_j-\delta_{ij}-2\nabla_{(i}v_{j)}-\nabla_i \nabla_j \ln m  )\,,
\end{equation}
%
yields a solution of the Einstein equations to leading order in $1/D$ describing a (possibly dynamical) configuration of a black brane horizon. Here $m(t,x^j)$ and $v^i(t,x^j)$ are the mass (and area) density and the velocity along the brane. Throughout this article we fix the units in the effective theory by normalizing the total (conserved) mass to $M=\int d^2 x\, m=1$.

MP black holes correspond to gaussian profiles for $m$, which become broader as their spin increases (a gaussian describes the area density of a large-dimensional sphere, see \cite{Andrade:2018nsz}). At low spin all the quasinormal modes of the solutions are stable, but when the spin reaches a critical value, a `bar-mode' instability appears, similar to those present in neutron stars, in which the horizon lengthens along one axis and shrinks along the transverse one. Ref.~\cite{Andrade:2018nsz} found an exact non-linear solution of \eqref{eqs} for a stationary rotating black bar (which, as we remarked, does not radiate when $D\to\infty$). It also identified zero-mode perturbations of black bars which are Gregory-Laflamme threshold modes of black strings when the bars are long.

\begin{figure}[t]
\centering
\includegraphics[width=.8 \linewidth]{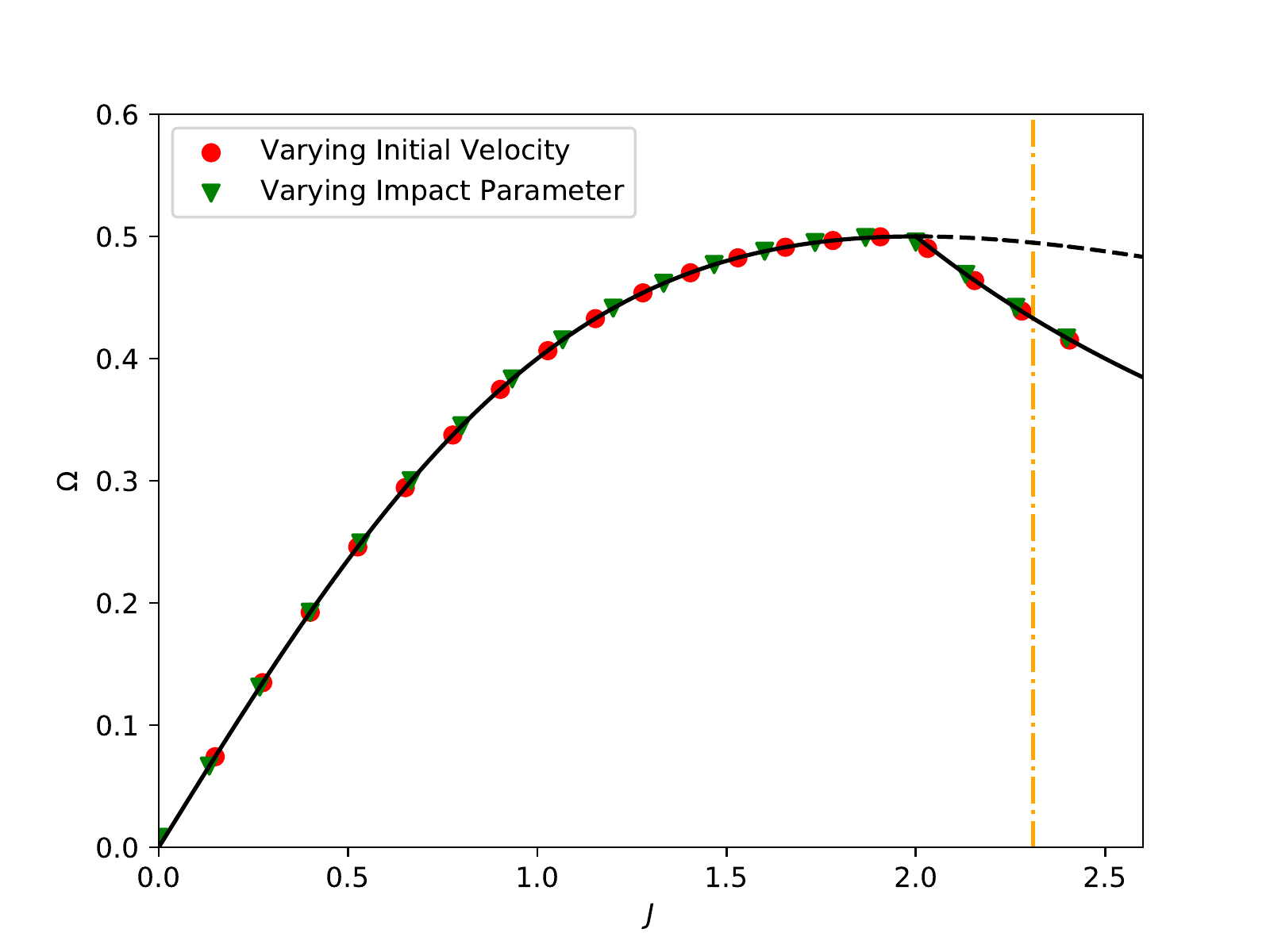}
\caption{Phase diagram for MP black holes and black bars \cite{Andrade:2018nsz} (dashed and continuous lines), and final states of the numerical evolution of a collision after $t = 10$ (circles and triangles). The vertical axis is the angular velocity of the horizon, the horizontal axis the total angular momentum for fixed mass.
The orange dash-dotted line at $J = 4/\sqrt{3}$ marks the stability limit for black bars.
For $J \gtrsim 2.43$ the intermediate bar configuration breaks, so we omit these values in this plot.}%
\label{fig:endpoint}
\end{figure}

Figure~\ref{fig:endpoint} shows these stationary phases in the plane of angular velocity $\Omega$ vs.\ angular momentum $J$. MP black holes exist for all $J$ but are bar-mode unstable for $J>2$, where black bars exist. 
With the effective-theory mass $M=1$, the dimensionless ratio of physical spin ${\bf J}$ to physical mass ${\bf M}$ is obtained as 
\beq\label{JM}
\frac{\mathbf{J}}{\mathbf{M}^\frac{D-2}{D-3}}
\simeq\frac{J}{\sqrt{2\pi e D}}
\eeq
(we set $16 \pi G=1$). Bear in mind that this expression is only asymptotically valid at large $D$, and not very accurate for moderate $D$, even for MP black holes. 

\medskip
\noindent\textbf{Colliding black holes at large $D$.} We set up two gaussian blobs (black holes) on a 2-brane in a collision course, and follow their evolution by numerically solving \eqref{eqs}. For simplicity we consider black holes with equal masses, zero spin, and equal but oppositely oriented velocities.

Before describing our results, let us review some aspects of black hole collisions at large $D$ \cite{Emparan:2013moa}. First, whenever $D>4$ it is well known that there are no bounded stable Keplerian orbits; hence there is no inspiral phase. The short range of the gravitational interaction at large $D$ then requires that the two black holes are aimed at one another with a sufficiently small impact parameter. This is not a problem for our purposes since we require $J$ of order one in \eqref{JM}, so the angular momenta and impact parameters are indeed small (in physical mass units). When the two black holes are within a distance $\ord{r_0/D}$ of each other (where $r_0$ is a characteristic length of the black hole), their horizons are expected to quickly merge on a time scale $\ord{r_0/D}$, presumably forming a tube of size $\ord{r_0/D}$ between them. At this point they will emit a burst of gravitational radiation of frequency $\omega=\ord{D/r_0}$.  The horizon then evolves quickly until it enters a regime of slower evolution, on time scales $\ord{r_0}$ controlled by \eqref{eqs}. 

When we evolve the entire collision using \eqref{eqs}, all the fast dynamics on time scales $\ord{r_0/D}$ is smoothed out. Since the thin black brane regulates the collision, there is properly never an instant in which the black holes merge, and the initial sharp outburst of high-frequency radiation is not visible. Missing this first part of the process is not an important shortcoming: we expect that this quick evolution is almost featureless, as indicated already by the universal nature of quasinormal modes of high frequencies $\ord{D/r_0}$ \cite{Emparan:2014cia,Emparan:2014aba}. The most interesting part of the dynamics is the one accurately captured by eqns.~\eqref{eqs} (which provide the non-linear theory of the ``featureful'' quasinormal modes of frequency  $\ord{1/r_0}$), \ie the evolution of the merged horizon.

We solve numerically the evolution equations by discretizing the spatial directions in a square domain with periodic boundary conditions. We have used two independent codes, with equivalent results: one is written in the {\sl Julia} language \cite{juliacode} and the other one in {\sl Mathematica}.
The {\sl Julia} code uses a two-dimensional Fourier grid with FFT differentiation in the spatial directions, and the {\sl DifferentialEquations.jl} package \cite{diffeqns} for time integration.
The {\sl Mathematica} code uses finite-difference differentiation in the spatial directions and a fourth-order Runge-Kutta method in the time direction.

\medskip
\noindent\textbf{Results.} We have performed numerical simulations of collisions with different initial velocities and impact parameters for the colliding black holes. Since there is no gravitational radiation, the total angular momentum $J$ is conserved throughout the evolution, which we have checked in our numerics. We have found that, for a large range of initial velocities and impact parameters, the value of $J$ is enough to predict the final state of the system, according to the stationary configurations that exist with that $J$: rotating MP black hole or black bar, stable or unstable. This is shown in Fig.~\ref{fig:endpoint}. 

That is, for collisions with $J < 2$ we obtain final states that correspond to MP black holes, which are the only stationary and stable phases in this range. For larger $J$ the MP black holes are unstable to bar formation and correspondingly we do not find these anymore as final states in our simulations. Instead, for $2 < J  < J_c=4/\sqrt{3}\approx 2.31 $ the final states are stable black bars. The critical value $J_c$ is given by the reflection-symmetric marginal 
mode of the bars found in \cite{Andrade:2018nsz}, which 
marks the beginning of the unstable region for the black bars that can be formed in our simulations.
For values of $J$ slightly larger than $J_c$ the bars are long-lived, and we do not observe their breaking in our numerics.  
However, for $J$ sufficiently high, the bars do split---more specifically, we observe this for $J \gtrsim 2.43$ for running times of order $t \sim 10$ (in units of $M=1$). These bars break after two turns or less, and the intermediate configurations can resemble more an evolving dumbbell, whose life-time decreases with $J$, than a quasi-stationary bar.

\begin{figure}[th]
\includegraphics[width=1. \linewidth]{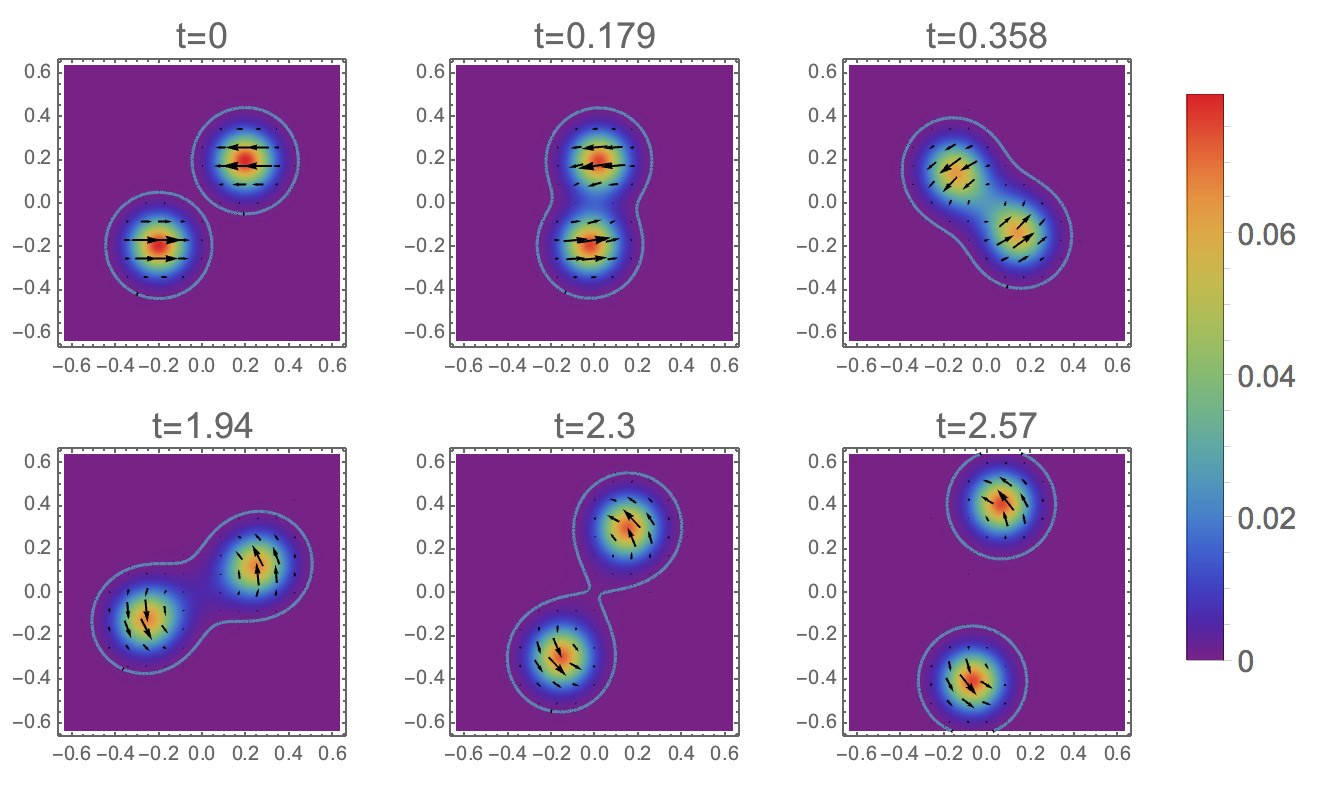}
\caption{Snapshots of the time evolution of a collision for $J = 2.43$. The density plots
show the energy density $m$, while the arrows depict $m v_i$. Contour lines corresponding to $m = 8 \times 10^{-4}$ 
are drawn to guide the eye.  After the black holes merge, they form a (deformed) bar that lasts for a time of 
order $\Delta t \approx 2.3$, until it breaks apart.}%
\label{fig:CCV}
\end{figure}

In Fig. \ref{fig:CCV} we show snapshots of the time evolution of a collision that yields CC violation. After the breaking, the two pieces of the bar fly apart and quickly settle into boosted MP black holes. Going to their rest frame, we observe that their approach to equilibrium is governed by the lowest-lying quasinormal mode computed in \cite{Andrade:2018nsz}. The final black holes have the same mass as the initial ones, and non-zero spin, but the total horizon area does not decrease in the process since when $D\to\infty$ the black hole area is not affected by its spin. Let us also note that  putting non-zero intrinsic spin on the initial black holes allows to demonstrate the formation of the long-lived bar and its subsequent instability in collisions with very small impact parameter \cite{longpaper}.

\medskip
\noindent\textbf{Discussion.} 
These results show that the evolution of the system for large enough $J$ forms a neck that quickly pinches down. Although our methods only allow to follow the evolution into regions of curvature smaller than $\ord{D}$, the evidence from \cite{Lehner:2010pn,Emparan:2018bmi}, and indeed what our simulations suggest, is that the horizon pinches off to zero size, leading to a violation of CC.

The combination of these lines of argument converges on the physical account described in the introduction. However, since we have worked in the leading order of the $1/D$ expansion, we must discuss effects that may arise at finite $D$, and the ensuing caveats for the picture we propose.

It is known that in the black string instability there exists a critical dimension $D_*\simeq 13.6$ \cite{Sorkin:2004qq,Emparan:2018bmi} such that when $D>D_*$, weakly non-uniform black strings are dynamically stable. However, this does not imply that non-uniform black bars become stable above a critical dimension. Metastable, slightly `bumpy' black bars for angular momenta in a narrow range above $J_c$ are consistent with the very long-lived phases that we observe for these spins. However, these are unlikely to persist at higher $J$; in contrast to black strings and black holes in a compact circle, there is nothing to hinder the centrifugally-driven separation between the two blobs that form in the instability and then fly apart. Moreover, it has been shown in \cite{Emparan:2018bmi} that in any finite $D$, the black string instability for thin enough black strings does not end in highly non-uniform black strings but rather proceeds towards pinch-off. In our collisions, black bars with large enough $J$ are analogous to these thin black strings and therefore we expect that they also evolve towards pinch off in any finite $D$ (provided radiation emission is subdominant, see below).
The centrifugal force will in fact drive the system more quickly towards pinch-off than in black strings.

It is therefore unclear whether the late-time fractal structure observed in the numerical evolutions of the instability in \cite{Lehner:2010pn,Figueras:2017zwa} has the time to develop in this system. Since the large-$D$ expansion and the ``thin black brane'' regulator that comes with it hide the details of the approach to the singularity, presumably numerical relativity simulations at finite $D$ are needed to address this question. Note, however, that our proposal for the breakup, namely, evaporation of a Planck-size neck, is actually independent of this, and we do not claim that the regulated large-$D$ evolution is evidence for it. Indeed, the presence of the regulator is irrelevant for the main outcome of our simulations, which is the formation of an intermediate bar-like configuration that becomes unstable. This convincingly shows that the system is driven towards a situation where CC will be violated, but the detailed features of the singularity and its formation are beyond the reach of our methods, and so are left to future studies.

Another caveat is that the size of the region of the horizon that is captured in the effective theory of \eqref{eqs} is only $\ord{1/\sqrt{D}}$. However, as shown very effectively in \cite{Emparan:2018bmi}, this range can be enlarged by incorporating $1/D$-corrections until the method reproduces accurately the detailed features of black holes and black strings at finite $D$ (including below the critical dimension). We do not see any apparent reason why \textit{perturbative} $1/D$ corrections should lead to qualitative, instead of merely quantitative, changes in our picture. 

More important are the consequences of \textit{non-perturbative} corrections in $1/D$. Of these, the loss of angular momentum through the emission of gravitational radiation is the main mechanism that opposes the instability: if the rate at which the angular momentum of the bar is shed off is faster than the instability growth, then the black bar may  spin down to a stable MP black hole before the neck has time to form and pinch down. We have used the $D$-dimensional quadrupole formula \cite{Cardoso:2002pa} to estimate the characteristic time scale $\tau_\text{rad}=\partial_t\ln (\mathbf{J/M})$ of the radiative spin-down of a rotating ellipsoidal bar. We have then compared it to the shortest characteristic time $\tau_\text{inst}$ for the growth of the Gregory-Laflamme instability of a black string of the same mass as the black bar. The radiation damping time is longer at large $D$ by a strong factor,
\beq
\frac{\tau_\text{rad}}{\tau_\text{inst}}\sim D^{D}\,,
\eeq
indeed so much so that our more accurate estimate suggests that already when $D\gtrapprox 7$ the spin-down may be much too slow to prevent the contraction of the neck \cite{longpaper}. 
%
Intermediate black bar states are in fact plausible only in $D\geq 6$, since only in these dimensions does the MP black hole have linear bar-mode instabilities \cite{Dias:2014eua} (however, see \cite{Shibata:2009ad}). These bar-modes have been followed non-linearly in $D=6,7,8$ and they return back  to a stable black hole through radiation emission \cite{Shibata:2010wz}. However, their long lifetime suggests that their angular momentum is not large enough to reach the unstable regime of black bars (moreover, the horizons in \cite{Shibata:2010wz} do not result from a collision merger). Our estimates are uncertain, but the suppression of radiation with increasing $D$ is so strong that we find it hard to envisage how the spin-down could be efficient in, say, $D\approx 10$ and possibly even lower dimensions. 
 
At any rate, although at present it is difficult to obtain a more precise estimate, we are confident that our analysis supports the conclusion that the violation of CC proposed here will be present in a high but finite $D$. 

\section*{Acknowledgments}

Work supported by ERC Advanced Grant GravBHs-692951 and MEC grant FPA2016-76005-C2-2-P. RL  is  supported  by  MEC grant  FPU15/01414.


\end{document}